\documentclass[aps,prd,twocolumn,a4paper,nofootinbib]{revtex4}
\usepackage{dcolumn}% Align table columns on decimal point
\usepackage{bm}% bold math
\usepackage{amssymb,amsmath}
\usepackage[english]{babel}
\usepackage[dvips]{graphicx}
\usepackage{subfigure}
\usepackage{color}
\usepackage{psfrag}

\def\lya{Ly$\alpha$}

\newcommand{\simgt}{\,\hbox{\lower0.6ex\hbox{$\sim$}\llap{\raise0.6ex\hbox{$>$}}}\,}
\newcommand{\simlt}{\,\hbox{\lower0.6ex\hbox{$\sim$}\llap{\raise0.6ex\hbox{$<$}}}\,}

\begin{document}
\thispagestyle{empty}

%\preprint{}

\title{Early Dark Energy at High Redshifts: Status and Perspectives}

\author{Jun-Qing Xia${}^1$}
\email{xia@sissa.it}
\author{Matteo Viel${}^{2,3}$}
\email{viel@oats.inaf.it}

\affiliation{${}^1$Scuola Internazionale Superiore di Studi
Avanzati, Via Beirut 2-4, I-34014 Trieste, Italy}

\affiliation{${}^2$INAF-Osservatorio Astronomico di Trieste, Via
G.B. Tiepolo 11, I-34131 Trieste, Italy}

\affiliation{${}^3$INFN/National Institute for Nuclear Physics, Via
Valerio 2, I-34127 Trieste, Italy}

\date{\today}

\begin{abstract}

Early dark energy models, for which the contribution to the dark
energy density at high redshifts is not negligible, influence the
growth of cosmic structures and could leave observable signatures
that are different from the standard cosmological constant cold dark
matter ($\Lambda$CDM) model.  In this paper, we present updated
constraints on early dark energy using geometrical and dynamical
probes. From WMAP five-year data, baryon acoustic oscillations and
type Ia supernovae luminosity distances, we obtain an upper limit of
the dark energy density at the last
scattering surface (lss), $\Omega_{\rm EDE}(z_{\rm
lss})<2.3\times10^{-2}$ ($95\%$ C.L.).  When we include higher
redshift observational probes, such as measurements of the linear
growth factors, Gamma-Ray Bursts (GRBs) and Lyman-$\alpha$ forest
(\lya), this limit improves significantly and becomes $\Omega_{\rm
EDE}(z_{\rm lss})<1.4\times10^{-3}$ ($95\%$ C.L.). Furthermore, we
find that future measurements, based on the Alcock-Paczy\'nski test
using the 21cm neutral hydrogen line, on GRBs and on the \lya\,
forest, could constrain the behavior of the dark energy component
and distinguish at a high confidence level between early dark energy
models and pure $\Lambda$CDM. In this case, the constraints on the
amount of early dark energy at the last scattering surface improve
by a factor ten, when compared to present constraints. We also
discuss the impact on the parameter $\gamma$, the growth rate index,
which describes the growth of structures in standard and in modified
gravity models.
\end{abstract}

%\pacs{98.80.Es, 95.36.+x, 98.80.-k}

\maketitle

%\numberwithin{equation}{section}

%%%%%%%%%%%%%%%%%%%%%%%%%%%%%%%%%%%%%%%%%%%%%%%%%%%%%%%%%%%%%%%%%
%%                 INTRODUCTION                                %%
%%%%%%%%%%%%%%%%%%%%%%%%%%%%%%%%%%%%%%%%%%%%%%%%%%%%%%%%%%%%%%%%%

\section{Introduction}
\label{intro}

Current cosmological observations, such as the cosmic microwave
background (CMB) measurements of temperature anisotropies and
polarization \cite{Komatsu:2008hk} and the redshift-distance
measurements of Type Ia Supernovae (SNIa) at $z<2$
\cite{Kowalski:2008ez}, have demonstrated that the Universe is now
undergoing an accelerated phase of expansion and that its total
energy budget is dominated by the dark energy component.  The nature
of dark energy is one of the biggest unsolved problems in modern physics and
has been extensively investigated in recent years, both under the theoretical
and the observational point of view. The simplest candidate of dark
energy is the cosmological constant ($\Lambda$CDM): however, this
model suffers of the fine-tuning and coincidence problems
\cite{CCproblem}.  To lift these tensions, many alternative
dynamical dark energy models, such as quintessence
\cite{Quintessence}, phantom \cite{Phantom}, K-essence
\cite{kessence} and quintom \cite{Feng:2004ad}, have been proposed
(for a review see Ref.\cite{copeland}).

Among all the dynamical dark energy models, we will focus here on
early dark energy (EDE) ones, in which a small fraction of dark energy
is present up to the last scattering surface (lss), unlike $\Lambda$CDM for
which $\Omega_{\rm DE}(z_{\rm lss})\simeq0$, The differences between
early dark energy models and pure $\Lambda$CDM are particularly
evident at high redshifts, over a large fraction of the cosmic time,
when the first structures form.  EDE has been shown to influence the
growth of cosmic structures (both in the linear and in the non-linear
regime), to change the age of the universe, to have an influence on
CMB physics, to impact on the reionization history of the
universe, to modify the statistics of giant arcs in strong cluster
lensing statistics
\cite{doran01,linderjenkins03,dolag04,mainini04,bartelmann06,fedeli07,crociani08,grossi08,francis08a}.

The reason for addressing this particular model of dark energy is
also driven by the increasing availability of high redshift
observations that somewhat bridge the gap between the CMB and very
local cosmological probes.  Dark energy studies rely mainly on: the
clustering properties of luminous red galaxies \cite{paddy},
baryonic acoustic oscillations (BAO)
\cite{Eisenstein:2005su,Cole:2005sx}, weak lensing data
\cite{li,wlother}, high redshift SNIa \cite{riess07}, the
\lya~forest \cite{McDonald:2004xn,lesgourgues07}, GRBs
\cite{Schaefer:2006pa,Li:2006ev} as standard candles, and also the
number of high redshift galaxies and clusters of galaxies
\cite{vikhlinin08} that could be studied with deep field
observations. Of course, since the present data sets seem to be in
good agreement with pure $\Lambda$CDM model, we hope that these
observations could open up a new high redshift window  on the
properties of dark energy and possible allow to confirm or disproof
its redshift evolution. Such a difficult measurement would have
profound implications for physical cosmology and particle physics.

In this paper we extend our studies of dark energy behavior into the
redshift range $2<z<1100$, which covers the so-called dark ages, and
present constraints on EDE from current and future cosmological
observations. We will use CMB, BAO, SNIa, \lya~forest observations
of growth factors and matter power spectrum and the GRBs from
current available data sets, while for forecasts we will add to
Planck-like observations, some higher redshift GRBs, some improved
\lya~constraints and a measurement that could be potentially very
important of the so-called Alcock-Paczy\'nski (AP, \cite{ap79}) test
using 21cm maps \cite{nusser05,barkana06}.  As we will see, the
particular parameterization of EDE chosen will be such that the
higher the redshift the tighter the constraints will be on the
parameters describing the given model.  Thereby, futuristic
intergalactic medium \lya~forest data, GRBs distance moduli and the
21cm maps, before or around reionization, are expected to be very
promising probes of dynamical dark energy models (see also
Ref.\cite{crociani08} for the impact of these models on the
reionization history of the universe). We note that the behavior of
dark energy in this redshift range has been exploited by
Ref.\cite{Mocker}, who constrained the EDE density of to be less
than $2\%$ of the total energy density. Here, we will improve this limit by
adding several different observations: as mentioned before this is
not only important to get stronger constraints on some cosmological
parameters but also to understand the consistency of different
probes and how systematic effects impact on the final derived measurements
(e.g. Refs.\cite{doransteffen07,linderrobbers08}).  Of course, we
stress that while our approach has the great advantage of using
dynamical and geometrical cosmological probes in the linear or
quasi-linear regime, other approaches, based for example on haloes
concentrations, could also be envisaged and must rely on an accurate
comparison with numerical simulations \cite{francis08a}.

Our paper is organized as follows: in Sec. II we describe the
theoretical framework of the early dark energy model and the datasets
we used. Sec. III contains our main global fitting results from the
current observations. In Sec. IV we present the forecasts for the
future measurements while Sec. V is dedicated to the conclusions and
discussion.

%%%%%%%%%%%%%%%%%%%%%%%%%%%%%%%%%%%%%%%%%%%%%%%%%%%%%%%%%%%%%%%%%
%%                  SECTION 1                                  %%
%%%%%%%%%%%%%%%%%%%%%%%%%%%%%%%%%%%%%%%%%%%%%%%%%%%%%%%%%%%%%%%%%

\section{Method and Data}

\subsection{Parameterization of early dark energy}

We decided to use the mocker model introduced in Ref.\cite{Mocker}
motivated by the following two observational facts: ${\rm i})$ the
best-fit model is pure $\Lambda$CDM model and the amount of dark
energy at the last scattering surface is constrained to be close to
zero from CMB and big bang nucleosynthesis (BBN) observations; ${\rm
 ii})$ an equation of state of the
dark energy component which is rapidly evolving (i.e. $dw/dz\gg1$) seems to
be ruled out by high-redshift SNIa \cite{riess07}.

For these EDE models the parameterization reads:
\begin{equation}
w_{\rm EDE}(a)=-1+\left[1-\frac{w_0}{1+w_0}a^C\right]^{-1}~,
\end{equation}
where $a=1/(1+z)$ is the scale factor, $w_0$ is the present
equation-of-state of dark energy and $C$ characterizes the ``running"
of the equation of state. Consequently, the evolution of dark energy
density can easily be obtained via energy conservation as:
\begin{equation}
\frac{\rho_{\rm EDE}(a)}{\rho_{\rm
EDE}(1)}=\left[(1+w_0)a^{-C}-w_0\right]^{3/C}~.
\end{equation}

\begin{figure}[t]
\begin{center}
\includegraphics[scale=0.45]{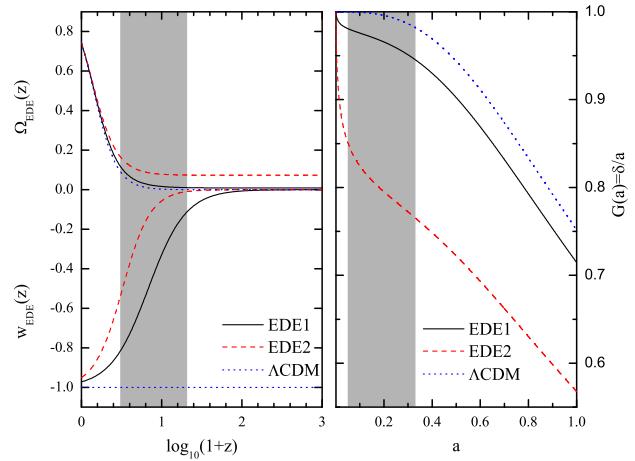}s
\caption{Left panel: the evolution of dark energy density and equation
  of state for three models, $w_0=-0.972$ and $C=1.858$ (EDE1),
  $w_0=-0.95$ and $C=2.5$ (EDE2) and pure $\Lambda$CDM. Right panel:
  the linear growth factor $G(a)=\delta(a)/a$ normalized at $z = \infty$
  for the three models.  The shaded area represents the redshifts that we
  will be mainly probing in the present work ($z\sim
  2-20$). \label{fig1}}
\end{center}
\end{figure}

In Fig.~\ref{fig1} we plot the dark energy density (upper part of
the left panel) and equation of state (bottom part of the left
panel) as a function of redshift for three different models: pure
$\Lambda$CDM (dotted blue line), EDE1 (solid black line) which has
$(w_0,C)=(-0.972,1.858)$ and EDE2 (dashed red line) which has
$(w_0,C)=(-0.95,2.5)$. All these models fit the CMB and the lower
redshift SNIa constraints very well. In the right panel, we also
show the linear growth rate (growth factors divided by the scale
factor) normalized at $z=\infty$ and compare the different
evolutions with redshifts in the three cases: differences of the
order of $20\%$ are visible at $z\sim 3$. In the two panels we also
show as a shaded vertical band the region in the redshift range
$z=2-20$, a period which is roughly $25\%$ of the age of the
universe. This is the redshift range we will be (mainly) focussing
on in the rest of the paper in order to discriminate between the
different dark energy models.

We stress that this is just one of the possible parameterizations
for early dark energy models, another one has been suggested by
Ref.\cite{wetterich04} and recently used in Ref.\cite{grossi08}.
However, we prefer to use the mocker model in order to compare with
the findings of Ref.\cite{Mocker} and because this parameterization
has a smooth redshift derivative at low $z$ for $w(z)$. We note that
one of the most important parameters is the amount of dark energy
during the structure formation period and this is given by ($a_{\rm
eq}$ is the matter-radiation equality scale factor):
\begin{equation}
{\Omega}_{\rm EDE,sf} = -(\ln\, a_{\rm eq})^{-1}\int^{0}_{\ln
          a_{\rm eq}} \Omega_{\rm EDE}\left(a\right){\rm d} \ln a \, ,
\label{eq:Omegasf}
\end{equation}
and we will also quote this value in the rest of the paper in order
to compare with other works and constraints as well (e.g.
Refs.\cite{doran01,fedeli07}).

We will also use some growth factors measurements obtained mainly
through \lya~forest observations and one lower redshift from Sloan
Digital Sky Survey (SDSS) galaxies. For this purpose we will compute
the growth factors for the EDE models with a subroutine implemented
in the publicly available Monte Carlo Markov Chains (MCMC) package
CosmoMC \cite{CosmoMC} that solves the second order differential
equations for the growth factor as in Ref.\cite{linderjenkins03}.
Furthermore, we also modify CosmoMC to include the perturbations of
dynamical dark energy models generally as done in Refs.\cite{pert}.

\subsection{Current and Future Datasets}
\subsubsection{Current Observations}

We will rely here on the following cosmological probes: ${\rm i})$
CMB anisotropies and polarization; ${\rm ii})$ baryonic acoustic
oscillations in the galaxy power spectra; ${\rm iii})$ SNIa distance
moduli; ${\rm iv})$ GRBs distance moduli; ${\rm v})$ the \lya~forest
growth factors and matter power spectrum measurements.

In the computation of CMB power spectra we have included the WMAP
five-year (WMAP5) temperature and polarization power spectra with
the routines for computing the likelihood supplied by the WMAP team
\cite{WMAP5:Other}. Besides the WMAP5 information, we also use some
distance-scale indicators.

BAOs have been detected in the current galaxy redshift survey data
from the SDSS and the Two-degree Field
Galaxy Redshift Survey (2dFGRS)
\cite{Eisenstein:2005su,Cole:2005sx,Huetsi:2005tp,BAO}. The BAO can
directly measure not only the angular diameter distance, $D_A(z)$,
but also the expansion rate of the universe, $H(z)$, which is
powerful for studying dark energy \cite{Albrecht:2006um}. Since
current BAO data are not accurate enough for extracting the
information of $D_A(z)$ and $H(z)$ separately \cite{Okumura:2007br},
one can only determine an effective distance
\cite{Eisenstein:2005su}:
\begin{equation}
D_v(z)\equiv\left[(1+z)^2D_A^2(z)\frac{cz}{H(z)}\right]^{1/3}~.
\end{equation}
In this paper we use the Gaussian priors on the distance ratios
$r_s(z_d)/D_v(z)$:
\begin{eqnarray}
r_s(z_d)/D_v(z=0.20)&=&0.1980\pm0.0058~,\nonumber\\
r_s(z_d)/D_v(z=0.35)&=&0.1094\pm0.0033~,
\end{eqnarray}
with a correlation coefficient of $0.39$, extracted from the SDSS
and 2dFGRS surveys \cite{BAO}, where $r_s$ is the comoving sound
horizon size and $z_d$ is drag epoch at which baryons were released
from photons given by Ref.\cite{Eisenstein:1997ik}.

The SNIa data provide the luminosity distance as a function of
redshift which is also a very powerful measurement of dark energy
evolution. The supernovae data we use in this paper are the recently
released Union compilation (307 samples) from the Supernova
Cosmology project \cite{Kowalski:2008ez}, which include the recent
samples of SNIa from the (Supernovae Legacy Survey) SNLS and ESSENCE
survey, as well as some older data sets, and span the redshift range
$0\lesssim{z}\lesssim1.55$. In the calculation of the likelihood
from SNIa we have marginalized over the nuisance parameter as done
in \cite{SNMethod} and ignored the systematic errors to improve our
results\footnote{In this paper, we find that the low redshifts
probes could not constrain on the early dark energy models very
well. Therefore, we focus on the constraints on the EDE model from
high redshifts probes. Furthermore, in Ref.\cite{xiaglobal} we
discussed the effect of constraints on dark energy models with or
without systematic error. When including the systematic error, the
constraints are slightly weaker, but the main results are not
changed.}.

Furthermore, we make use of the Hubble Space Telescope (HST)
measurement of the Hubble parameter $H_{0}\equiv
100\,h$~km~s$^{-1}$~Mpc$^{-1}$ by a Gaussian likelihood function
centered around $h=0.72$ and with a standard deviation $\sigma=0.08$
\cite{HST}.

In order to constrain the early dark energy model at high redshifts,
we also include the GRBs, linear growth factors and \lya~forest
data.

GRBs can potentially be used to measure the luminosity distance out to
higher redshift than SNIa. Recently, several empirical correlations
between GRB observables were reported, and these findings have
triggered intensive studies on the possibility of using GRBs as
cosmological ``standard'' candles. However, due to the lack of
low-redshift long GRBs data to calibrate these relations, in a
cosmology-independent way, the parameters of the reported correlations are
given assuming an input cosmology and obviously depend on the same
cosmological parameters that we would like to constrain. Thus,
applying such relations to constrain cosmological parameters leads to
biased results. In Ref.\cite{xiagrb} the circular problem is naturally
eliminated by marginalizing over the free parameters involved in the
correlations; in addition, some results show that these correlations do
not change significantly for a wide range of cosmological parameters
\cite{Firmani}. Therefore, in this paper we use the 69 GRBs sample
over a redshift range from $z=0.17-6.60$ published in
Ref.\cite{Schaefer:2006pa} but we keep in mind the issues related to
the ``circular problem" that are more extensively discussed in
Ref.\cite{xiagrb}.

In Table I we list two types of linear growth factors data we use:
${\rm i})$ the normalization $\sigma_8$ inferred from the SDSS
\lya~power spectrum\footnote{These data at different redshifts are
not totally independent. The derived $\sigma_8$ values are obtained
by using the covariance matrix as measured from the data of the SDSS
flux power spectrum and there are (very weak) correlations in the
flux power between different redshift bins. So, although not totally
independent, we regard the derived values of the power spectrum
amplitude from the SDSS flux power as nearly independent.}
\cite{Viel:2004bf,Viel:2005ha}; ${\rm ii})$ the linear growth rate
$f\equiv\Omega^\gamma_{\rm m}$ from galaxy power spectrum at low
redshift \cite{Hawkins:2002sg} and \lya~growth factor measurement
obtained from the SDSS \lya~power spectrum at $z=3$ by
Ref.\cite{McDonald:2004xn}.

\begin{table}
TABLE I. $\sigma_8$ values and growth rates with $1\sigma$ error
bars from which we derived the linear growth factors used in our
analysis.
\begin{center}

\begin{tabular}{cccc}

\hline

~~~~~$z$~~~~ & ~~~~$\sigma_8$~~~~ & ~~~$\sigma_{\sigma_8}$~~~& ~~Ref.~~~\\

\hline

2.125&0.95&0.17&\cite{Viel:2004bf}\\
2.72&0.92&0.17&\\

\hline

2.2&0.92&0.16&\cite{Viel:2005ha}\\
2.4&0.89&0.11&\\
2.6&0.98&0.13&\\
2.8&1.02&0.09&\\
3.0&0.94&0.08&\\
3.2&0.88&0.09&\\
3.4&0.87&0.12&\\
3.6&0.95&0.16&\\
3.8&0.90&0.17&\\

\hline

$z$ & $f$ & $\sigma_{f}$& Ref.\\

\hline

0.15&0.51&0.11&\cite{Hawkins:2002sg}\\
3.00&1.46&0.29&\cite{McDonald:2004xn}\\

\hline
\end{tabular}
\end{center}
\end{table}

We have used two \lya~forest data sets: ${\rm i})$ the high resolution
QSO absorption spectra presented in Ref.\cite{Viel:2004bf} consisting
of the LUQAS sample \cite{Kim:2003qt} and the reanalyzed data in
Ref.\cite{Croft:2000hs} (C02, this data set as a whole will be
labelled as VHS in the following); ${\rm ii})$ the SDSS \lya~forest
sample presented in Ref.\cite{McDonald:2004eu}.  The SDSS \lya~forest
data set consists of $3035$ QSO spectra with low resolution ($R\sim
2000$) and low S/N ($< 10$ per pixel) spanning a wide range of
redshifts ($z=2.2-4.2$), while the LUQAS and the C02 samples contain
57 high resolution ($R \sim 45000$), high signal-to-noise ($>50$ per
pixel) QSO spectra with median redshifts of $z=2.125$ and $z=2.72$,
respectively.  The flux power spectrum of the \lya~forest is the
quantity which is observed and needs to be modeled at the percent or
sub-percent level using accurate numerical simulations that
incorporate the relevant cosmological and astrophysical processes, in
order to extract the underlying (linear) dark matter power spectrum.
In this paper, we will use the derived linear power spectrum measured
by the data set of VHS, based on ``VHS\lya'' together with the SDSS
\lya~power spectrum ``SDSS\lya'', since they are in agreement and this
latter has a stronger constraining power.  More precisely, the
VHS\lya~power spectrum consists of estimates of the linear dark matter
power spectrum at nine values in the wavenumber space $k$ at $z=2.125$
and nine values at $z=2.72$, in the range $0.003<k$(s/km)$<0.03$,
while the SDSS\lya~consists of a single measurement at $z=3$ and
$k=0.009$ s/km of amplitude, slope and curvature. The estimate of the
uncertainty of the overall amplitude of the matter power spectrum is
$29\%$ for the first dataset and $\sim 14\%$ for the second. This
estimate takes into account possible systematic and statistical errors
(see the relevant tables of VHS for a detailed discussion). The code
assigns a Gaussian prior to the corresponding nuisance parameter and
marginalize over it. Finally, we will also rely on the growth
factor measurements from the two data sets, derived using numerical
simulations of the observed flux power, and we will label this as GF\lya.

\subsubsection{Future Measurements}
\label{secfut}

In order to forecast future measurements we will use the same
observables as before without BAO but exploiting the role of 21cm
maps at high redshifts.

For the simulation with PLANCK \cite{Planck}, we follow the method
given in Ref.\cite{xiaplanck} and mock the CMB temperature (TT) and
polarization (EE) power spectra and temperature-polarization cross
correlation (TE) by assuming a given fiducial cosmological
model. In Table II, we list the assumed experimental specifications
of the future (mock) Planck measurement.

\begin{table}
TABLE II. Assumed experimental specifications for the mock Planck-like
measurements. The noise parameters $\Delta_T$ and $\Delta_P$ are given
in units of $\mu$K-arcmin.
\begin{center}
\begin{tabular}{cccccc}
\hline \hline

$f_{\rm sky}$~ & ~$l_{\rm max}$~ & (GHz) &
~$\theta_{\rm fwhm}$~ & ~$\Delta_T$~~ & ~~$\Delta_P$~ \\

\hline

 0.65 & 2500 & 100 & 9.5' & 6.8 & 10.9 \\
      &      & 143 & 7.1' & 6.0 & 11.4 \\
      &      & 217 & 5.0' & 13.1 & 26.7 \\

\hline \hline
\end{tabular}
\end{center}
\end{table}

The proposed satellite SNAP (Supernova / Acceleration Probe) will
be a space based telescope with a one square degree field of view
that will survey the whole sky \cite{SNAP}. It aims at increasing
the discovery rate of SNIa to about $2000$ per year in the redshift
range $0.2<z<1.7$. In this paper we simulate about $2000$ SNIa
according to the forecast distribution of the SNAP
\cite{Kim:2003mq}. For the error, we follow the
Ref.\cite{Kim:2003mq} which takes the magnitude dispersion to be
$0.15$ and the systematic error $\sigma_{\rm sys}=0.02\times z/1.7$.
The whole error for each data is given by $\sigma_{\rm
mag}(z_i)=\sqrt{\sigma^2_{\rm sys}(z_i)+0.15^2/{n_i}}~$, where $n_i$
is the number of supernovae of the $i'$th redshift bin. Furthermore,
we add as an external data set a mock set of 400 GRBs, that mimic SWIFT \cite{swift} observations, in the
redshift range $0 < z < 6.4$ with an intrinsic dispersion in the
distance modulus of $\sigma_{\mu}=0.16$ and with a redshift
distribution very similar to that of Figure 1 of
Ref.\cite{Hooper:2005xx}.

For the linear growth factors data, we simulate the mock data from
the fiducial model of Table I with the error bars reduced by a
factor of two. This is probably reasonable given the larger amounts
of \lya~forest data (e.g. SDSS-III or the X-shooter spectrograph
\cite{Xshooter}) that will become available soon as long with a
better control of several systematics errors and more importantly on
the thermal history of the intergalactic medium.

We also simulate the \lya~power spectrum, consisting of three data
points at $z=2, 3$ and 3.5 and at wavenumbers of 0.002 s/km, 0.009
s/km and 0.02 s/km, with $1\%$ fractional errors in a similar way as
done by Ref.\cite{Gratton:2007tb}, in which a forecasting on the
perspective of constraining the neutrino masses was presented.

The availability of 21cm (1420 MHz) maps with future experiments
like LOFAR \cite{lofar} will allow to measure in a given frequency
interval and for a given position in the sky the differential
brightness temperature of a patch of gas against the CMB. Thus a
completely new window on the neutral hydrogen content of the
universe, regarded as a tracer of cosmic structures and of
astrophysical processes, will be opened soon. The AP test
\cite{ap79} assumes that the for an isotropically distributed set of
astronomical objects the separations along and across the
line-of-sights scale differently and thus generate anisotropies. The
geometrical distortion depends on the quantity $[H(z)\,D_A(z)]^{-1}$
that thereby can be measured and provide constraints on some
parameters. The extended LOFAR will operate between 30 and 240 MHz,
with spatial resolution of the order of arcseconds, probing the
universe in the redshift range $z\sim 6-45$. For our purposes here
we will assume that the quantity $[H(z)\,D_A(z)]^{-1}$, will be
measured at $z=6.5,~8,~12$ with $5\%$ precision. Although other
measurements could be performed using 21cm datasets, such as (quite
optimistically) a full measurement of the matter power spectrum at
very high redshifts, we will concentrate here only on the AP test,
since this appears to be easier to be performed and it is less
sensitive to the astrophysical uncertainties that relate the
differential brightness temperature to the matter density (see
however Ref.\cite{geil08} for other possible and very promising
constraints). It is nevertheless important to note that LOFAR sensitivity
drops for $z>12$ and useful HI maps at these redshifts could only be
obtained with future ground based telescopes like SKA \cite{ska} and
with a better understanding of the foregrounds that contaminate the
cosmological signal (see for example the discussion in
Ref.\cite{nusser05}).

Here, we will use a somewhat more conservative approach and simulate
a measurement of the quantity described above in a redshift range
accessible to the LOFAR experiment. Although challenging, we believe
that the AP test could be performed and give constraints similar to
those obtained here.

%%%%%%%%%%%%%%%%%%%%%%%%%%%%%%%%%%%%%%%%%%%%%%%%%%%%%%%%%%%%%%%%%
%%                  RESULTS                                    %%
%%%%%%%%%%%%%%%%%%%%%%%%%%%%%%%%%%%%%%%%%%%%%%%%%%%%%%%%%%%%%%%%%
\section{Results from present datasets}
\label{results}

\begin{table*}
TABLE III. Constraints on the early dark energy model from the current
observations. Here we show the mean and the best fit values. For
some parameters that are only weakly constrained we quote the $95\%$
upper limit.
\begin{center}

\begin{tabular}{|c|c|c|c|c|c|c|}

\hline

Parameter& \multicolumn{2}{c|}{WMAP5+BAO+SN} & \multicolumn{2}{c|}{+GRB+\lya} & \multicolumn{2}{c|}{+GRB+GF\lya}\\

\cline{2-7}

& Mean & BestFit &Mean&BestFit&Mean&BestFit\\

\hline

$w_0$ & $<-0.906$ & $-0.972$ & $<-0.952$ & $-0.999$ & $<-0.958$ & $-0.997$\\
$C$ & $<2.711$ & $1.858$ & $<2.613$ & $1.628$ & $<2.245$ & $0.152$\\
$\Omega_m$ & $0.261\pm0.014$ & $0.258$ & $0.283\pm0.014$ & $0.279$ & $0.275\pm0.014$ & $0.274$\\
$\sigma_8$ & $0.748\pm0.049$ & $0.734$ & $0.842\pm0.022$ & $0.863$ & $0.836\pm0.024$ & $0.846$\\
$\Omega_{\rm EDE}(z_{\rm lss})$ & $<0.0228$ & $0.0064$ & $<0.0029$ & $1.77\times10^{-6}$ & $<0.0014$ & $1.71\times10^{-9}$\\
${\Omega}_{\rm{EDE,sf}}$&$0.0643\pm0.0076$&$0.0672$&$0.0540\pm0.0029$&$0.0529$&$0.0546\pm0.0024$&$0.0543$\\
$\gamma$ & $-$ & $-$ & $0.622\pm0.139$ & $0.552$ & $-$ & $-$\\

\hline
\end{tabular}
\end{center}
\end{table*}

In our analysis, we perform a global fitting using the CosmoMC
package.  We assume purely adiabatic initial conditions and a flat
universe, with no tensor contribution. We vary the following
cosmological parameters with top-hat priors: the dark matter energy
density $\Omega_{c} h^2 \in [0.01,0.99]$, the baryon energy density
$\Omega_{b} h^2 \in [0.005,0.1]$, the primordial spectral index $n_s
\in [0.5,1.5]$, the primordial amplitude $\log[10^{10} A_s] \in
    [2.7,4.0]$ and the angular diameter of the sound horizon at last
    scattering $\theta \in [0.5,10]$. For the pivot scale we set
    $k_{s0}=0.05\,$Mpc$^{-1}$. When CMB data are included, we also
    vary the optical depth to reionization $\tau \in [0.01,0.8]$. We
    do not consider any massive neutrino contribution. From the
    parameters above the MCMC code derives the reduced Hubble
    parameter $H_0$, the present matter fraction $\Omega_{{\rm m}0}$,
    $\sigma_8$, and $\Omega_{\rm EDE,sf}$: so, these parameters have
    non-flat priors and the corresponding bounds must be interpreted
    with some care. In addition, CosmoMC imposes a weak prior on the
    Hubble parameter: $h \in [0.4,1.0]$. For part of the analysis, we
    also present constraints on the growth index $\gamma$, defined as:
\begin{equation}
\frac{d\ln \delta(a)}{d \ln a}=\Omega_{\rm m}(a)^{\gamma}\; .
\end{equation}
This parameter, introduced by \cite{wangstein98}, provides a good
fit to the growth rate for many different cosmological models and
depends on the growth of perturbations, $H(z)$ and $\Omega_{\rm
{m}0}$. It has been recently measured by
Refs.\cite{nesseris08,diporto08} using mainly the observed growth
factors at high and low redshifts and values different from that
expected for the pure $\Lambda$CDM model ($\gamma \sim 0.55$) could
be important for constraining both quintessence models and modified
gravity scenarios (see Ref.\cite{gammapapers,zhao} for forecasting
and a more extensive discussion on linear growth rate in modified
gravity models).

\subsection{CMB+BAO+SNIa}

\begin{figure}[t]
\begin{center}
\includegraphics[scale=0.45]{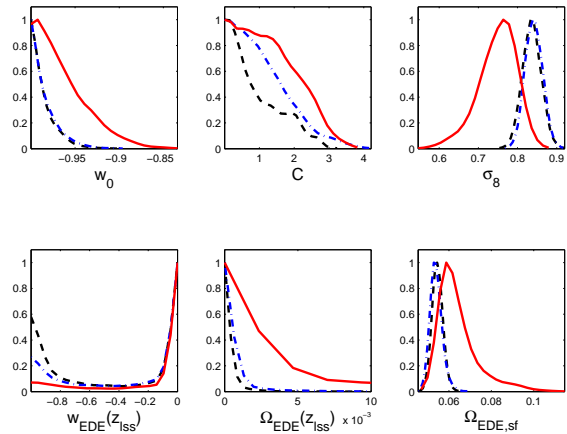}
\caption{1D current (marginalized) constraints on the dark energy
parameters $w_0$, $C$, as well as $\sigma_8$, $w_{\rm EDE}(z_{\rm
lss})$, $\Omega_{\rm EDE}(z_{\rm lss})$ and ${\Omega}_{\rm EDE,sf}$,
from different current data combinations: WMAP5+BAO+SN (red solid
lines), WMAP5+BAO+SN+GRB+\lya~(blue dash-dot lines) and
WMAP5+BAO+SN+GRB+GF\lya~(black dashed lines).\label{fig2}}
\end{center}
\end{figure}

In Table III we show our main global constraints on the early dark
energy model from different current data combinations.

Firstly, we present the constraints from the data combination: CMB+
BAO+SNIa. In Fig.~\ref{fig2} we show the one dimensional
distributions of some cosmological parameters from this data
combination (solid red lines). For the parameters describing the
equation of state of early dark energy, the constraints are still
weak, namely the $95\%$ upper limits are $w_0<-0.906$ and $C<2.711$.
The limit on the present equation of state of dark energy component
is still consistent with other recent results in the literature
\cite{Komatsu:2008hk,xiaglobal}. However, current observational
data, could not determine the ``running" of equation of state very
well. The reason is that the data we use are only the low redshift
BAO and SNIa ($z<2$), as well as the very high redshift CMB data at
last scattering surface ($z\sim1090$). Due to the lack of the
cosmological probes in the dark ages ($2<z<1100$), we do not know
exactly the evolution of equation of state of early dark energy.
Therefore, the constraint on the ``running" of equation of state,
which is parameterized by $C$, is rather poor.

Consequently,  current observations still allow very large amount of
early dark energy at high redshift $z\sim1090$ as $\Omega_{\rm
EDE}(z_{\rm lss})<0.0228$ ($95\%$ C.L.), which is consistent with
the results obtained by Refs.\cite{Mocker,Lee:2002nv}. Early dark
energy models with a non-negligible fraction of dark energy density
still fit the data very well. Furthermore, there are two peaks in
the posterior distribution of equation of state of early dark energy
at CMB last scattering surface $w_{\rm EDE}(z_{\rm lss})$. The first
peak around $w_{\rm EDE}(z_{\rm lss})=0$ is in fact caused by the
poor constraint on $C$. When $C$ is larger than one, the equation of
state $w_{\rm EDE}$ will approach zero at high redshift. The larger
$C$ is, the faster $w_{\rm EDE}$ approaches $w=0$ (we will come back
to this point in Sect. IV).  Considering the weak constraint on $C$,
there is significant probability that allows $w_{\rm EDE}(z_{\rm
lss})=0$. The second peak around $w_{\rm EDE}(z_{\rm lss})=-1$ can
be understood easily since the pure $\Lambda$CDM model is still
favored by the current observations. When $w_0$ is close to $-1$ and
$C$ is close to $0$, which is consistent with the data, $w_{\rm
EDE}(z_{\rm lss})$ reaches the cosmological constant value.
Therefore, in order to remove the first peak, we need a much more
stringent constraint on $C$ from future measurements.

\begin{figure}[t]
\begin{center}
\includegraphics[scale=0.45]{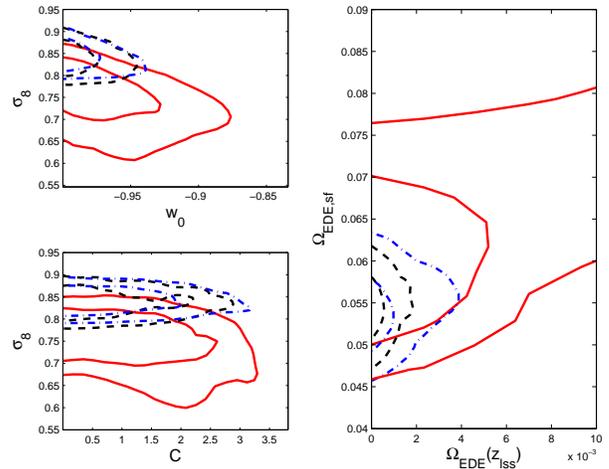}
\caption{2D (marginalized) contours of some parameters from
different current data combinations: WMAP5+BAO+SN (red solid lines),
WMAP5+BAO+SN+GRB+\lya~(blue dash-dot lines) and
WMAP5+BAO+SN+GRB+GF\lya~(black dashed lines).\label{fig3}}
\end{center}
\end{figure}

As we showed in Sect. II, early dark energy models also affect structure
formation \cite{doran01}. From Fig.~\ref{fig1} we can see that the
linear growth factor of early dark energy model is suppressed
significantly by the large value of parameter $C$. The
non-negligible dark energy density at high redshifts slows down the
linear growth function of the matter perturbation and  leads to a
low value of $\sigma_8$ today. Using CMB+BAO+SNIa data combination,
illustrated in Fig.~\ref{fig2}, we obtain the limit on $\sigma_8$
today of $\sigma_8=0.748\pm0.049$ ($68\%$ C.L.), which is obviously
lower than one obtained in the pure $\Lambda$CDM framework:
$\sigma_8=0.812\pm0.026$ \cite{Komatsu:2008hk}, while the error bar
is enlarged by a factor of two. Therefore, the $\sigma_8$ today and
$C$ are anti-correlated, as shown in Fig.~\ref{fig3}. On the other
hand, the $\sigma_8$ and equation of state $w_0$ at present also
have an anti-correlation. When $w_0$ is increased, the fraction of
dark energy density becomes large, which also leads the lower
$\sigma_8$ today.

\begin{figure}[t]
\begin{center}
\includegraphics[scale=0.45]{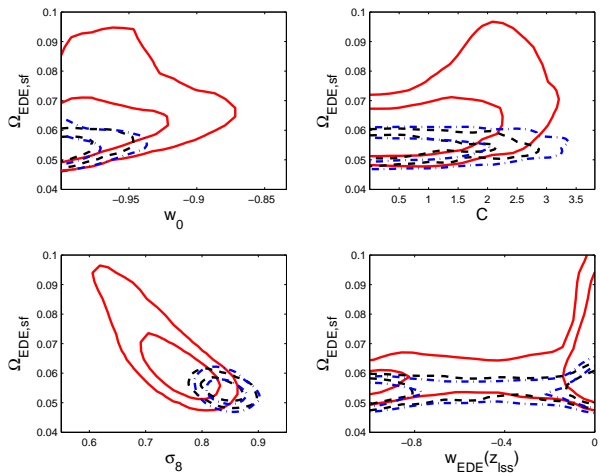}
\caption{2D (marginalized) contours of ${\Omega}_{\rm EDE,sf}$
versus $w_0$, $C$, $\sigma_8$ and $w_{\rm EDE}(z_{\rm lss})$ from
different current data combinations: WMAP5+BAO+SN (red solid lines),
WMAP5+BAO+SN+GRB+\lya~(blue dash-dot lines) and
WMAP5+BAO+SN+GRB+GF\lya~(black dashed lines).\label{fig4}}
\end{center}
\end{figure}

Finally, we discuss the average value of the fraction of dark energy
in the structure formation era ${\Omega}_{\rm EDE,sf}$. In Table III
we can find that the current constraint is ${\Omega}_{\rm
EDE,sf}=0.0643\pm0.0076$ at $1\sigma$ confidence level. From
Eq.(\ref{eq:Omegasf}) and Fig.~\ref{fig3} it is easy to see that a
large $\Omega_{\rm EDE}(z_{\rm lss})$ leads to the high
${\Omega}_{\rm EDE,sf}$. In Fig.~\ref{fig4} we also plot the two
dimensional plots between ${\Omega}_{\rm EDE,sf}$ and other related
parameters. As we discussed before, when $w_0$ and $C$ become large,
the fraction of dark energy density at high redshifts will also be
large. On the other hand, when ${\Omega}_{\rm EDE,sf}$ increases,
the linear growth function of matter perturbation will be slowed
down and the value of $\sigma_8$ today will decrease consequently:
${\Omega}_{\rm EDE,sf}$ and the $\sigma_8$ today are then obviously
anti-correlated.

To sum up, using CMB, BAO and SNIa, we constrain the amount of dark
energy at the last scattering surface to be less than 2\% and in the
structure formation era to be of the order of 6\%. This last number is
in rough agreement with the results of \cite{doran06}, who used WMAP 3
year data and some external data sets, although their values are
somewhat smaller ($\sim 4\%$) and this is due to the slightly different
definition of ${\Omega}_{\rm EDE,sf}$.

\subsection{Adding High-Redshift Probes}

From the results above, we can see that without observations in or
around the so called dark ages, the constraints on the early dark
energy model are interesting but not very tight. In this subsection
we add some high-redshift observational data, such as GRBs, \lya~and
GF\lya~data, to improve our constraints. Because most of GF\lya~data
are obtained from the \lya~forest power spectrum, in order to avoid
overestimating the weight of \lya~data, in our analysis we do not use GF\lya~and
\lya~data together. And the results, as we expect, from these two
data combinations are consistent between each other. Therefore, in
the following we mainly present the global fitting results obtained
from those that include GRB and GF\lya~data.

When adding the high-redshift probes, the constraints on $w_0$ and
$C$ improve significantly: $w_0<-0.958$ and $C<2.245$ at $95\%$
confidence level, while their best fit values are very close to the
pure $\Lambda$CDM model. Naturally, the $95\%$ upper limit of
$\Omega_{\rm EDE}(z_{\rm lss})$ has also been tightened by a factor
of ten, $\Omega_{\rm EDE}(z_{\rm lss})<1.4\times10^{-3}$. A large
number of early dark energy models have been ruled out by the
high-redshift probes.  The high-redshift observations are indeed
effective in constraining further the early dark energy models
investigated here. Note that adding only the power spectrum
information as provided by \lya~improves the constraints on
$\Omega_{\rm EDE}(z_{\rm lss})$ by a factor five, when compared to
CMB+BAO+SNIa, so the growth factor information is really fundamental
in constraining the EDE model. However, the double peak in the
distribution of $w_{\rm EDE}(z_{\rm lss})$ is still present, which
implies that the current data are not accurate enough to measure
$w_{\rm EDE}(z_{\rm lss})$, even when high-redshift probes are
added.

Another effect of adding high-redshift probes is the larger value of
$\sigma_8$ today. As we know, the \lya~forest data favor a larger
$\sigma_8\sim 0.9$ than the CMB measurement\footnote{This fact does
not imply a contradiction or a failure of the $\Lambda$CDM model but
there is indeed a tension between small scale constraints and the
large scale ones. However, more recent studies that combine WMAP
year 3, the Lyman-$\alpha$ forest and weak-lensing measurements of
the COSMOS-z survey \cite{lesgourgues07} point to a value of
$\sigma_8=0.800\pm 0.023$, which is in good agreement with the
findings of WMAP year 5 plus BAO and SNe \cite{Komatsu:2008hk}.}.
When we add the GF\lya~or \lya~forest data, we obtain a higher
$\sigma_8$ today than before: $\sigma_8=0.836\pm0.024$ ($1\sigma$).
However, this is still smaller than the value favored by \lya~forest
data alone in the pure $\Lambda$CDM model and consistent with
WMAP5+BAO+SN measurement of \cite{Komatsu:2008hk}. We also note that
if we use only VHS\lya~ data sets and adding them to WMAP5+BAO+SN
and GRBs the results do not change significantly as compared to
having SDSS\lya~and VHS\lya~together. This is mainly due to the
constraining power of the growth factors measurement obtained with
VHS\lya.

The conclusion of this section is that adding higher redshift probes
constrains the energy density of dark energy at the last scattering
surface to be around $0.1\%$.

\subsection{Growth Index $\gamma$}
\label{secgamma}

Finally, we extend our discussion by briefly addressing the
perspective of constraining the linear growth index $\gamma$.
Measurements of the growth history of cosmic structures combine
information on both cosmic expansion and the underlying theory of
gravity, the parameter $\gamma$ is thus a unique prediction of any
modified gravity model and a simultaneous fitting of this parameter
and other dynamical probes could provide constraints on modified
gravity scenarios and dynamical dark energy models.

By using all the current data available we obtain
$\gamma=0.622\pm0.139\,$ (1$\sigma$) error bar for our EDE model,
which is in agreement with the $\Lambda$CDM and with the values
obtained by Refs.\cite{nesseris08,diporto08} (with slightly smaller
error bars). The present error on $\gamma$ is also similar to that
inferred from future weak lensing and SNIa data by
Refs.\cite{gammapapers}.

\begin{figure}[t]
\begin{center}
\includegraphics[scale=0.45]{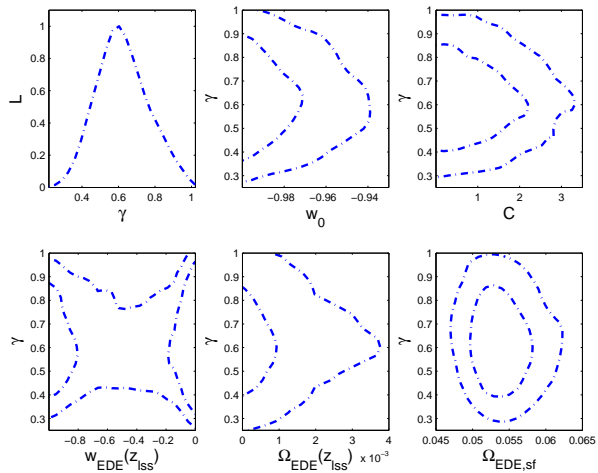}
\caption{The posterior distribution of $\gamma$ and 2D
(marginalized) contours of $\gamma$ and $w_0$, $C$, $w_{\rm
EDE}(z_{\rm lss})$, $\Omega_{\rm EDE}(z_{\rm lss})$ and
${\Omega}_{\rm EDE,sf}$ from WMAP5+BAO+SN+GRB+\lya~(blue dotted
lines).\label{fig5}}
\end{center}
\end{figure}

In Figure \ref{fig5} we show the 2-dimensional contours of $\gamma$
vs. $\sigma_8,C,w_0$ and $\Omega_{\rm EDE}(z_{\rm lss})$ to address
possible degeneracies of this parameter. For the current data sets,
we note that the degeneracies are weak and only the degeneracy in
the $\gamma-{\Omega}_{\rm EDE,sf}$ plane is significant. This is
easily understood since a larger $\gamma$ implies a faster growth of
structures and thereby a smaller amount of dark energy.

\section{Results from future data sets}

\begin{table}
TABLE IV. Constraints on the early dark energy model from future
measurements. We also list the standard deviation of these
parameters based on the future mock data sets. For parameters that
are only weakly constrained  we quote the $95\%$ upper limit
instead.
\begin{center}

\begin{tabular}{|c|c|c|}

\hline

Parameter&Fiducial $\Lambda$CDM & Fiducial EDE \\

\hline

$w_0$& $<-0.975$ & $[-0.985,-0.920]$\\
$C$& $<2.635$ & $[0.256,2.300]$\\
$\Omega_m$& $0.0029$ & $0.0032$\\
$\sigma_8$& $0.0078$ & $0.0145$\\
$\Omega_{\rm EDE}(z_{\rm lss})$& $<1.02\times10^{-4}$ & $<0.011$\\
${\Omega}_{\rm{EDE,sf}}$&$0.0011$&$0.0037$\\
$\gamma$ & $0.026$ & $0.048$\\

\hline
\end{tabular}
\end{center}
\end{table}

\begin{figure}[t]
\begin{center}
\includegraphics[scale=0.45]{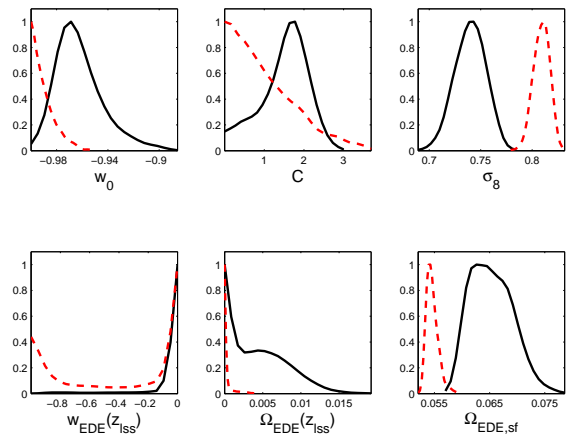}
\caption{1D current (marginalized) constraints on the dark energy
parameters $w_0$, $C$, as well as $\sigma_8$, $w_{\rm EDE}(z_{\rm
lss})$, $\Omega_{\rm EDE}(z_{\rm lss})$ and ${\Omega}_{\rm EDE,sf}$
from future measurements with the fiducial models: $\Lambda$CDM (red
dashed lines), EDE1 model (black solid lines).\label{fig6}}
\end{center}
\end{figure}

Since the present data do not give very stringent constraints on the
parameters of early dark energy model, it is worthwhile discussing
whether future data could determine these parameters conclusively.
For this purpose we have performed a further analysis and we have
chosen two fiducial models in perfect agreement with current data: a
pure $\Lambda$CDM model and an EDE model with parameters taken to be
the best-fit values of Table III from the current constraints of
CMB+BAO+SNIa. This latter model will be labelled EDE1, and its
evolution is shown in Fig.\ref{fig1}.

\subsection{Fiducial $\Lambda$CDM}

Firstly, we choose the pure $\Lambda$CDM as the fiducial model. In
Table IV we list the forecasts for some parameters using all the
future measurements described in Section \ref{secfut}: GRBs, SNIa,
\lya~(matter power spectrum and growth factors) and the AP test.

\begin{figure}[t]
\begin{center}
\includegraphics[scale=0.45]{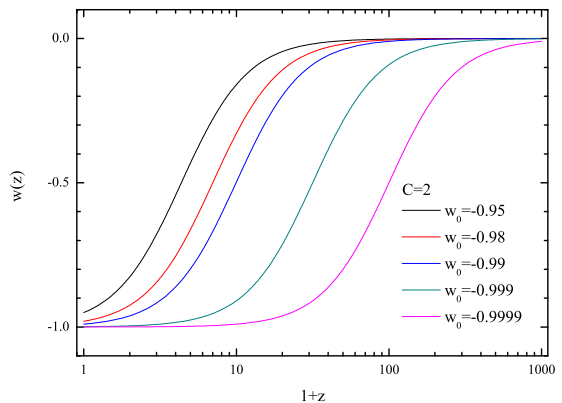}
\caption{Evolution of different equation of state for the Mocker
model when fixing $C=2$. All these models fit low redshift ($z<2$)
and CMB constraints but have very different evolution at
intermediate redshifts. \label{fig7}}
\end{center}
\end{figure}

Due to the smaller error bars of the mock data sets, the constraint
on $w_0$ improves significantly, namely the $95\%$ upper limit is
now $w_0<-0.975$. The fraction of early dark energy density has been
limited very stringent, $\Omega_{\rm EDE}(z_{\rm
lss})<1.02\times10^{-4}$ ($95\%$ C.L.), which is reduced by another
factor of ten, compared with the present constraint. The error on
the $\gamma$ parameter that we forecast is $0.026$, while the error
on the amount of EDE in the structure formation era is about
$1.1\times 10^{-3}$. Therefore, we can foresee a very precise
determination of the amount of dark energy in the dark ages (or in
the structure  formation era) with this future data set.

By contrast, the $95\%$ upper limit on $C$ has not been improved,
$C<2.635$. This is due to the chosen parameterization of early dark
energy. In Fig.~\ref{fig7} we illustrate the different evolution of
several EDE equations of state varying $w_0$ and keeping $C=2$. We
note that the smaller the redshift at which $w_0$ approaches the
cosmological constant, the smaller the redshift at which $w(z)$
flattens to zero in the high redshift universe. For example, if
$w_0=-0.95$, we have $w(z=10)\sim-0.1$ and $C=2$ has been ruled out
obviously. On the other hand, if $w_0=-0.9999$, we still have
$w(z=10)\sim-1$. Consequently, $C=2$ is still favored by the data.
Therefore, even if the constraint on $w_0$ becomes stringent, a
large value of $C$ is still allowed, and then the constraint on $C$
will be somewhat poorer. Moreover, due to the allowed large value of
$C$, the first peak of $w_{\rm EDE}(z_{\rm lss})$ around $w=0$ is
still present, as shown in Fig.~\ref{fig6}, where the 1D marginalized
constraints at the last scattering surface are shown.

\subsection{Fiducial EDE}

\begin{figure}[t]
\begin{center}
\includegraphics[scale=0.4]{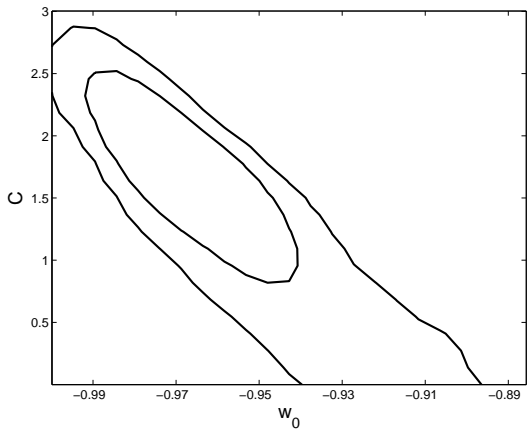}
\caption{2D marginalized contours $C$ vs $w_0$ from future mock
measurements of the fiducial EDE1 model.\label{fig8}}
\end{center}
\end{figure}

For our EDE fiducial model we choose the best fit values of current
constraint as the fiducial model to simulate the future
measurements, $w_0=-0.972$ and $C=1.858$ (EDE1).

In Fig.~\ref{fig6}, we plot the one dimensional distributions of
$w_0$ and $C$. The $95\%$ confidence level is $-0.985<w_0<-0.920$
and $0.256<C<2.300$, respectively. These results imply that the
future measurements could distinguish the pure $\Lambda$CDM model
and early dark energy models at least at $5\sigma$ confidence level.
We also show the contour in the ($w_0$,$C$) plane in
Fig.~\ref{fig8}. We note that even with these futuristic data sets
simulated in a conservative way the value $C=0$ is still allowed at
a $2\sigma$ confidence level. When $w_0$ increases, $C$ must
decrease in order to match the current observations. Thus, $w_0$ and
$C$ are anti-correlated. Consequently, the allowed value of
$\Omega_{\rm EDE}(z_{\rm lss})$ could be very large: $\Omega_{\rm
EDE}(z_{\rm lss})<1.1\times10^{-2}$ ($95\%$ C.L.). Furthermore, with
this fiducial model, the double peak in the distribution of $w_{\rm
EDE}(z_{\rm lss})$ disappears: because the pure $\Lambda$CDM is
disfavored at more than $5\sigma$ confidence level.

In Fig.~\ref{fig6} we clearly appreciate the very distinctive
predictions of ${\Omega}_{\rm EDE,sf}$ and $\sigma_8$ for the
$\Lambda$CDM and EDE1 models. The values obtained for ${\Omega}_{\rm
EDE,sf}$ and $\sigma_8$ are different by $20\%$ and $10\%$,
respectively, and can be measured at very high precision in the two
cases.

We are also interested in understanding which of the different
cosmological probes give the most stringent constraints. In order to
investigate this we perform three different runs: we find that the
best constraints are obtained when future \lya~and GF\lya~are added to
Planck and SNAP data, while adding AP only to Planck and SNAP results
in the the worst constraints. In the third run, we consider adding
only GRBs mock data to Planck and SNAP and in this case the results
are in between those inferred from \lya~and AP test. The degeneracies
between different parameters are similar in three cases. We also
checked that degrading the accuracy of the AP test to 10\% as opposed
to the chosen 5\% has not a significant impact on the recovered
parameters.

We also consider the results in terms of $\gamma$ has done in
Section \ref{secgamma}. We find that we could constrain $\gamma$ to
high significance in the two cases: the standard deviations are
$0.026$ and $0.048$, which are reduced by a factor of five and three, when
compared to the current constraints, respectively. By looking at
Fig. 9, we can now see that the constraints on $\gamma$ are more
precise and degeneracies with $\Omega_{\rm EDE}$($z_{\rm lss}$) and
$\Omega_{\rm EDE,sf}$ are stronger.

\begin{figure}[t]
\begin{center}
\includegraphics[scale=0.45]{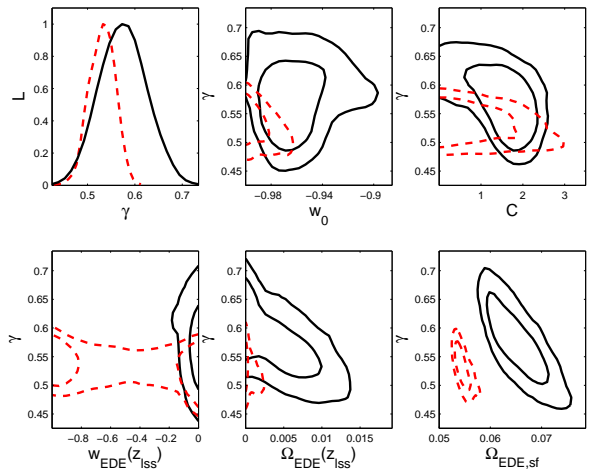}
\caption{The posterior distribution of $\gamma$ and 2D
(marginalized) contours of $\gamma$ and $w_0$, $C$, $w_{\rm
EDE}(z_{\rm lss})$, $\Omega_{\rm EDE}(z_{\rm lss})$ and
${\Omega}_{\rm EDE,sf}$ from future measurements with the fiducial
models: $\Lambda$CDM (red dashed lines), EDE1 model (black solid
lines).\label{fig9}}
\end{center}
\end{figure}

\section{Conclusions and discussion}

With the availability of new data from SNIa, and the advent of
relatively new cosmological probes, such as \lya~and GRBs in the
high redshift universe, there has been an increasing interest in the
study of dynamical dark energy models with a significant energy
contribution in the structure formation era. In this paper we
present constraints on a particular early dark energy model using
the latest observations. We use the Mocker parameterization proposed
by Ref.\cite{Mocker}, but our results are general since the amount
of dark energy at the last scattering surface and in the structure
formation era have also been quoted in order to better compare
with other possible dark energy parameterizations \cite{doran06}.

We find that current constraints based on CMB, BAO and SNIa data
limit the amount of dark energy density at the last scattering
surface to be $2\%$, while in the structure formation period is of
the order of $6\%$. If we add high redshift GRBs and the \lya~forest
data, in the form of a matter power spectrum or in the form of a
growth factor measurement, the constraints improve by a factor five
and fifteen, respectively. The $\Lambda$CDM model is still a very
good fit to the data currently available and only upper limits on
the amount of dark energy at the last scattering could be quoted.

We also forecast what could be done with future data sets in
particular: Planck for the CMB; GRBs and SNIa luminosity distances
as could be measured by SNAP; a better measurement of \lya~matter
power spectrum and growth factors as could be obtained from SDSS-III
and the AP test performed on HI 21 cm maps from LOFAR. With these
futuristic mock data sets we simulate two different fiducial models
that are currently in agreement with observations a $\Lambda$CDM
model and an EDE model with about $6\%$ contribution to the energy
density in the structure formation epoch. We find that the two
models will be clearly distinguishable at more than $5\sigma$
confidence level. Thereby, the role of these high redshift probes is
crucial in discriminating between pure $\Lambda$CDM and quintessence
and/or modified gravity models.

Furthermore, motivated by the fact that the growth factors at high
redshift seem to have the highest constraining power, we also
address the perspectives of measuring the parameter $\gamma$ that
parameterizes the growth rate of structures and is usually a very
precise prediction for any quintessence and/or modified gravity
model. Even in this case, we find that in the two models we could
constrain $\gamma$ to high significance: the standard deviations
are reduced by a factor of five and three, when compared with the current
constraint, respectively.

We conclude this Section with some words of caution: while high
redshift probes are important in constraining early dark energy
models, the obtained values from present and future constraints rely mainly on the \lya~forest and
its derived growth factors. These measurements, summarized partly in
Table I, have been obtained in the framework of pure $\Lambda$CDM
model and are valid only when small deviations from this model are
considered. We believe that we are in such a situation here, where
the amount of dark energy has little impact on observed
\lya~quantities and this effect is possibly degenerate with other
parameters that have been already marginalized over to obtain the
final estimate.  However, further work using both numerical
hydrodynamical simulations as long with an accurate control of systematics should be done in order to better
check this (see for example Ref.\cite{delya}). Along the same lines,
while the role of SNIa as cosmological probes is somewhat reasonably
well established, those of GRBs is still controversial. We tried to
circumvent this problem by marginalizing over the nuisance
parameters that enter the calibration of the GRBs distance moduli
but, even in this case, more observational and theoretical work is
needed in order to trust quantitatively the results. Even regarding
the forecasting using the AP test, we tried to implement these
constraints in a conservative way, in a range of redshift that could
be observed by LOFAR, without assuming that a full matter power
spectrum could be measured from the observed brightness temperature
(this of course will have a fundamental impact especially at even
higher redshifts than those considered here).

However, even considering these cautious remarks, our work clearly
emphasizes the capabilities of present and future data in an
intermediate redshift range which is poorly probed by the
observations, to quantitatively constrain the dark energy component
and its evolution. We consider these observations as very important
since they can open up a completely new window on observational
cosmology and high energy physics and provide new insights on either
modified gravity or quintessence scenarios.

\section*{ACKNOWLEDGMENTS}
The numerical analysis has been performed on the Shanghai
Supercomputer Center (SSC). We thank Viviana Acquaviva for useful
discussions.

%\svspace{\stretch{1}}

%\newpage


\begin{thebibliography}{50}


\bibitem{Komatsu:2008hk}
E.~Komatsu {\it et al.}, arXiv:0803.0547.

\bibitem{Kowalski:2008ez}
M.~Kowalski {\it et al.}, arXiv:0804.4142.

\bibitem{CCproblem}
S.~Weinberg, Rev.\ Mod.\ Phys.\ {\bf 61}, 1 (1989); I.~Zlatev,
L.~M.~Wang and P.~J.~Steinhardt, Phys.\ Rev.\ Lett.\ {\bf 82}, 896
(1999).

\bibitem{Quintessence}
R.~D.~Peccei, J.~Sola and C.~Wetterich, Phys.\ Lett.\ B {\bf 195},
183 (1987); C.~Wetterich, Nucl.\ Phys.\ B {\bf 302}, 668 (1988);
B.~Ratra and P.~J.~E.~Peebles, Phys.\ Rev.\ D {\bf 37}, 3406 (1988);
P.~J.~E.~Peebles and B.~Ratra, Astrophys.\ J.\ {\bf 325}, L17
(1988).

\bibitem{Phantom}
R.~R.~Caldwell, Phys.\ Lett.\ B {\bf 545}, 23 (2002).

\bibitem{kessence}
T.~Chiba, T.~Okabe and M.~Yamaguchi, Phys.\ Rev.\ D {\bf 62}, 023511
(2000); C.~Armendariz-Picon, V.~F.~Mukhanov and P.~J.~Steinhardt,
Phys.\ Rev.\ Lett.\ {\bf 85}, 4438 (2000).

\bibitem{Feng:2004ad}
B.~Feng, X.~L.~Wang and X.~Zhang, Phys.\ Lett.\ B {\bf 607}, 35
(2005).

\bibitem{copeland}
E.~J.~Copeland, M.~Sami and S.~Tsujikawa, Int.\ J.\ Mod.\ Phys.\  D
{\bf 15}, 1753 (2006).

\bibitem{doran01}
M.~Doran, J.~M.~Schwindt and C.~Wetterich, Phys.\ Rev.\  D {\bf 64},
123520 (2001).

\bibitem{linderjenkins03}
E.~V.~Linder and A.~Jenkins, Mon.\ Not.\ Roy.\ Astron.\ Soc.\  {\bf
346}, 573 (2003).

\bibitem{dolag04}
K.~Dolag, M.~Bartelmann, F.~Perrotta, C.~Baccigalupi, L.~Moscardini,
M.~Meneghetti and G.~Tormen, Astron.\ Astrophys.\  {\bf 416}, 853
(2004).

\bibitem{mainini04}
R.~Mainini, A.~V.~Maccio, S.~A.~Bonometto and A.~Klypin, Astrophys.\
J.\  {\bf 599}, 24 (2003).

\bibitem{crociani08}
D.~Crociani, M.~Viel, L.~Moscardini, M.~Bartelmann and
M.~Meneghetti, Mon.\ Not.\ Roy.\ Astron.\ Soc.\ {\bf 385}, 728
(2008).

\bibitem{bartelmann06}
M.~Bartelmann, M.~Doran and C.~Wetterich, Astron.\ Astrophys.\  {\bf
475}, 27 (2006).

\bibitem{fedeli07}
C.~Fedeli and M.~Bartelmann, Astron.\ Astrophys.\  {\bf 461}, 49
(2007).

\bibitem{francis08a}
M.~Manera and D.~F.~Mota, Mon.\ Not.\ Roy.\ Astron.\ Soc.\  {\bf
371}, 1373 (2006); M.~J.~Francis, G.~F.~Lewis and E.~V.~Linder,
arXiv:0810.0039; D.~F.~Mota, JCAP {\bf 0809}, 006 (2008).

\bibitem{grossi08}
M.~Grossi and V.~Springel, arXiv:0809.3404.

\bibitem{paddy}
N.~Padmanabhan {\it et al.}, Mon.\ Not.\ Roy.\ Astron.\ Soc.\  {\bf
378}, 852 (2007).

\bibitem{Eisenstein:2005su}
D.~J.~Eisenstein {\it et al.} Astrophys.\ J.\  {\bf 633}, 560
(2005).

\bibitem{Cole:2005sx}
S.~Cole {\it et al.} Mon.\ Not.\ Roy.\ Astron.\ Soc.\  {\bf 362},
505 (2005).

\bibitem{wlother}
Y.~Gong, T.~J.~Zhang, T.~Lan and X.~L.~Chen, arXiv:0810.3572;
M.~Kilbinger {\it et al.}, arXiv:0810.5129.

\bibitem{li}
H.~Li {\it et al.}, arXiv:0812.1672.

\bibitem{riess07}
A.~G.~Riess {\it et al.}, Astrophys.\ J.\  {\bf 659}, 98 (2007).

\bibitem{lesgourgues07}
J.~Lesgourgues, M.~Viel, M.~G.~Haehnelt and R.~Massey, JCAP {\bf
0711}, 008 (2007).

\bibitem{McDonald:2004xn}
P.~McDonald {\it et al.}, Astrophys.\ J.\  {\bf 635}, 761 (2005).

\bibitem{Schaefer:2006pa}
B.~E.~Schaefer, Astrophys.\ J.\  {\bf 660}, 16 (2007).

\bibitem{Li:2006ev}
H.~Li, M.~Su, Z.~Fan, Z.~Dai and X.~Zhang, Phys.\ Lett.\  B {\bf
658}, 95 (2008).


\bibitem{vikhlinin08}
A.~Vikhlinin {\it et al.}, arXiv:0812.2720.

\bibitem{ap79}
C.~Alcock and B.~Paczy\'nski, Nature {\bf 281}, 358 (1979).

\bibitem{nusser05}
A.~Nusser, Mon.\ Not.\ Roy.\ Astron.\ Soc.\  {\bf 364}, 743 (2005).

\bibitem{barkana06}
R.~Barkana, Mon.\ Not.\ Roy.\ Astron.\ Soc.\  {\bf 372}, 259 (2006).

\bibitem{Mocker}
E.~V.~Linder, Astropart.\ Phys.\ {\bf 26}, 16 (2006).

\bibitem{linderrobbers08}
E.~V.~Linder and G.~Robbers, JCAP {\bf 0806}, 004 (2008).

\bibitem{doransteffen07}
M.~Doran, S.~Stern and E.~Thommes, JCAP {\bf 0704}, 015 (2007).

\bibitem{wetterich04}
C.~Wetterich, Phys.\ Lett.\  B {\bf 594}, 17 (2004).

\bibitem{CosmoMC}
A.~Lewis and S.~Bridle, Phys.\ Rev.\ D {\bf 66}, 103511 (2002); URL:
http://cosmologist.info/cosmomc/.

\bibitem{pert}
J.~Weller and A.~M.~Lewis, Mon.\ Not.\ Roy.\ Astron.\ Soc.\  {\bf
346}, 987 (2003); G.~B.~Zhao, J.~Q.~Xia, M.~Li, B.~Feng and
X.~Zhang, Phys.\ Rev.\  D {\bf 72}, 123515 (2005); J.~Q.~Xia,
G.~B.~Zhao, B.~Feng, H.~Li and X.~Zhang, Phys.\ Rev.\  D {\bf 73},
063521 (2006).

\bibitem{WMAP5:Other}
J.~Dunkley {\it et al.}, arXiv:0803.0586; E.~L.~Wright {\it et al.},
arXiv:0803.0577; M.~R.~Nolta {\it et al.}, arXiv:0803.0593; B.~Gold
{\it et al.}, arXiv:0803.0715; G.~Hinshaw {\it et al.},
arXiv:0803.0732; URL: http://lambda.gsfc.nasa.gov/.

\bibitem{Huetsi:2005tp}
G.~Huetsi, Astron.\ Astrophys.\  {\bf 449}, 891 (2006).

\bibitem{BAO}
W.~J.~Percival {\it et al.}, Mon.\ Not.\ Roy.\ Astron.\ Soc.\ {\bf
381}, 1053 (2007).

\bibitem{Albrecht:2006um}
A.~Albrecht {\it et al.}, arXiv:astro-ph/0609591.

\bibitem{Okumura:2007br}
T.~Okumura, T.~Matsubara, D.~J.~Eisenstein, I.~Kayo, C.~Hikage,
A.~S.~Szalay and D.~P.~Schneider, Astrophys.\ J.\  {\bf 677}, 889
(2008).

\bibitem{Eisenstein:1997ik}
D.~J.~Eisenstein and W.~Hu, Astrophys.\ J.\  {\bf 496}, 605 (1998).

\bibitem{SNMethod}
E.~Di~Pietro and J.~F.~Claeskens, Mon.\ Not.\ Roy.\ Astron.\ Soc.\
{\bf 341}, 1299 (2003).

\bibitem{xiaglobal}
J.~Q.~Xia, H.~Li, G.~B.~Zhao and X.~Zhang, Phys.\ Rev.\ D {\bf 78},
083524 (2008).

\bibitem{HST}
W.~L.~Freedman {\it et al.}, Astrophys.\ J.\ {\bf 553}, 47 (2001).

\bibitem{xiagrb}
H.~Li, J.~Q.~Xia, J.~Liu, G.~B.~Zhao, Z.~H.~Fan and X.~Zhang,
Astrophys.\ J.\  {\bf 680}, 92 (2008).

\bibitem{Firmani}
C.~Firmani, V.~Avila-Reese, G.~Ghisellini and G.~Ghirlanda, Rev.\
Mex.\ Astron.\ Astrofis.\ {\bf 43}, 203 (2007).

\bibitem{Viel:2004bf}
M.~Viel, M.~G.~Haehnelt and V.~Springel, Mon.\ Not.\ Roy.\ Astron.\
Soc.\  {\bf 354}, 684 (2004).

\bibitem{Viel:2005ha}
M.~Viel and M.~G.~Haehnelt, Mon.\ Not.\ Roy.\ Astron.\ Soc.\  {\bf
365}, 231 (2006).

\bibitem{Hawkins:2002sg}
E.~Hawkins {\it et al.}, Mon.\ Not.\ Roy.\ Astron.\ Soc.\  {\bf
346}, 78 (2003); L.~Verde {\it et al.}, Mon.\ Not.\ Roy.\ Astron.\
Soc.\  {\bf 335}, 432 (2002).

\bibitem{Kim:2003qt}
T.~S.~Kim, M.~Viel, M.~G.~Haehnelt, R.~F.~Carswell and S.~Cristiani,
Mon.\ Not.\ Roy.\ Astron.\ Soc.\  {\bf 347}, 355 (2004).

\bibitem{Croft:2000hs}
R.~A.~C.~Croft {\it et al.}, Astrophys.\ J.\  {\bf 581}, 20 (2002).

\bibitem{McDonald:2004eu}
P.~McDonald {\it et al.}, Astrophys.\ J.\ Suppl.\  {\bf 163}, 80
(2006).

\bibitem{Planck}
Planck Collaboration, arXiv:astro-ph/0604069.

\bibitem{xiaplanck}
J.~Q.~Xia, H.~Li, G.~B.~Zhao and X.~Zhang, Int.\ J.\ Mod.\ Phys.\  D
{\bf 17}, 2025 (2008).

\bibitem{SNAP}
URL: http://snap.lbl.gov/.

\bibitem{Kim:2003mq}
A.~G.~Kim, E.~V.~Linder, R.~Miquel and N.~Mostek, Mon.\ Not.\ Roy.\
Astron.\ Soc.\  {\bf 347}, 909 (2004).

\bibitem{Hooper:2005xx}
D.~Hooper and S.~Dodelson, Astropart.\ Phys.\  {\bf 27}, 113 (2007).

\bibitem{swift} URL: http://heasarc.gsfc.nasa.gov/docs/swift/swiftsc.html

\bibitem{Xshooter}
URL: http://www.eso.org/instruments/xshooter/index.html/.

\bibitem{Gratton:2007tb}
S.~Gratton, A.~Lewis and G.~Efstathiou, Phys.\ Rev.\  D {\bf 77},
083507 (2008).

\bibitem{lofar}
URL: http://www.lofar.org/.

\bibitem{geil08}
S.~Wyithe, A.~Loeb and P.~Geil, Mon.\ Not.\ Roy.\ Astron.\ Soc.\
{\bf 383}, 1195 (2008).

\bibitem{ska}
URL: http://www.skatelescope.org/.

\bibitem{wangstein98}
P.~J.~E.~Peebles, {\it Large-Scale Structure of the Universe},
(Princeton U. Press: 1980); L.~M.~Wang and P.~J.~Steinhardt,
Astrophys.\ J.\  {\bf 508}, 483 (1998).

\bibitem{nesseris08}
S.~Nesseris and L.~Perivolaropoulos, Phys.\ Rev.\  D {\bf 77},
023504 (2008).

\bibitem{diporto08}
C.~Di Porto and L.~Amendola, Phys.\ Rev.\  D {\bf 77}, 083508
(2008).

\bibitem{gammapapers}
D.~Huterer and E.~V.~Linder,  Phys.\ Rev.\  D {\bf 75}, 023519
(2007); E.~V.~Linder and R.~N.~Cahn, Astropart.\ Phys.\ {\bf 28},
481, (2007).

\bibitem{zhao}
G.~B.~Zhao, L.~Pogosian, A.~Silvestri and J.~Zylberberg,
arXiv:0809.3791; V.~Acquaviva, A.~Hajian, D.~N.~Spergel and S.~Das,
Phys.\ Rev.\ D {\bf 78}, 043514 (2008).

\bibitem{Lee:2002nv}
W.~Lee and K.~W.~Ng, Phys.\ Rev.\  D {\bf 67}, 107302 (2003).

\bibitem{doran06}
M.~Doran and G.~Robbers, JCAP {\bf 0606}, 026 (2006).

\bibitem{delya}
M.~Viel, S.~Matarrese, T.~Theuns, D.~Munshi and Y.~Wang, Mon.\ Not.\
Roy.\ Astron.\ Soc.\  {\bf 340}, L47 (2003); P.~McDonald and
D.~J.~Eisenstein, Phys.\ Rev.\ D {\bf 76}, 063009 (2007).

\end{thebibliography}
\end{document}